\documentstyle[twoside,fleqn,espcrc2,epsfig,amssymb]{article}

\newcommand{\AmS}{{\protect\the\textfont2
  A\kern-.1667em\lower.5ex\hbox{M}\kern-.125emS}}

\title{New results from APE with nonperturbatively improved Wilson fermions\thanks
{Talk given by D. Becirevic at ``Lattice98'', Boulder, CO. He acknowledges the partial 
supports of: IN2P3, CNRS, and the organizers of the conference.}}

\author{D.~Becirevic$^{\rm a}$, Ph.~Boucaud\address{Universit\'e de Paris Sud, L.P.T.H.E.~(B\^at.~211), 
91405 Orsay-Cedex, France}, L.~Giusti\address{Scuola Normale Superiore and INFN, P.zza dei Cavalieri 7, 
I-56100 Pisa, Italy}, J.P.~Leroy$^{\rm a}$, V.~Lubicz\address{Dip. di Fisica, Univ. di Roma Tre and INFN, 
Via della Vasca Navale 84, I-00146 Rome, Italy}, G.~Martinelli$^{\rm d}$, F.~Mescia$^{\rm d}$, F.~Rapuano
\address{Dip. di Fisica, Univ. ``La Sapienza" and INFN, P.le A. Moro, I-00185 Rome, Italy}}       

\begin{document}
\newcommand{\sze}{\small}
\newcommand{\bea}{\begin{eqnarray}}
\newcommand{\eea}{\end{eqnarray}}
 
\begin{abstract}
We present the results for hadronic spectrum, decay constants and the light quark masses obtained with 
non-perturbatively improved Wilson fermions. We also give our preliminary results for the heavy-light decay constants.
\end{abstract}
 
\maketitle

The method for systematic improvement of the Wilson action and bilinear quark operators, recently proposed 
by {\sc Alpha}--collaboration \cite{alpha}, provides a full cancellation of the linear discretization errors. 
In other words, the results of lattice calculations of any 
$on-shell$ quantity, do converge to the continuum limit ($a\to 0$) much faster , {\it i.e.} with the rate ${\cal{O}}(a^2)$.  

To verify and quantify the effect of this improvement ({\sl first improved studies of the spectroscopy were done 
in}~\cite{alpe}), we generated 100 independent gauge field configurations on a $24^3\times 64$ lattice, 
at $\beta=6.2$, in the quenched approximation. We used the non-perturbative(NP) value of the coefficient for the 
improvement of the action, $c_{_{SW}}$, which guarantees the full improvement of the hadron mass spectrum. 
The values of all other counterterms coefficients, as well as of the renormalization constants which we used in 
this work, are listed in Tab.~\ref{tab:tt1}. Note that the NP results for $Z_S$, $Z_P$ and $Z_T$ (given in 
Tab.~\ref{tab:tt1} at $\mu a={1}$), are new.
We calculated several correlators for mesons made 
of $four$ degenerate quark `flavors'. In the analysis of all hadronic quantities (spectrum and decay constants), 
we used the method of `physical lattice planes', proposed and extensively used in Ref.~\cite{allton}. Further 
details on the computation and the analysis 
can be found in \cite{light}. Here, we will address several issues which are 
relevant for the improvement, and report our results.

\begin{table}[t]
\setlength{\tabcolsep}{0.3pc}
\caption{\sl \footnotesize Renormalization constants $Z_J$, and the improvement coefficients $c_J$ and $b_J$ at 
$\beta=6.2$ ($c_{_{SW}}=1.614$). For the constants which are not computed non-perturbatively ($\times$), we take 
the result of the 1-$loop$ boosted perturbation theory (BPT).
\vspace*{1mm}}
\label{tab:tt1}
\hspace*{-2mm}
\vspace*{-6mm}
\begin{tabular*}
{7.825cm}{|c|cc|cc|cc|}
\hline
\vspace*{-3.5mm}&&&&&&\\
\sze {${\it{J}}$ } &\sze $Z_J^{^{BPT}}$ &\sze $Z_J^{^{NP}}$ &\sze $b_J^{^{BPT}}$ & \sze $b_J^{^{NP}}$&\sze 
 $c_J^{^{BPT}}$ &\sze  $c_J^{^{NP}}$ \\ \hline \hline
\vspace*{-3mm}&&&&&&\\
\sze {${\it{S}}$ } & \sze 0.66 & \sze {\bf 0.61(1)$^\bullet$}& \sze 1.30 & \sze {$\times$}& \sze -- & \sze --  \\
\vspace*{-2.5mm}&&&&&&\\
\sze {${\it{V}}$ } &\sze  0.85 &\sze  {\bf 0.79}&\sze 1.24 &\sze  {\bf 1.40} & \sze -0.026 &\sze {\bf -0.21(7)} \\
\vspace*{-2.5mm}&&&&&&\\
{$\sze {\it{T}}$ } &\sze  0.93 & \sze {\bf 0.85(1)$^\bullet$}& \sze 1.22 &\sze {$\times$}& \sze 0.018&\sze {$\times
$}  \\ 
\vspace*{-2.5mm}&&&&&&\\
{$\sze {\it{A}}$ } &\sze  0.86 & \sze {\bf 0.81}& \sze 1.24 & \sze {$\times$} & \sze -0.012 &\sze {\bf -0.037}  \\
\vspace*{-2.5mm}&&&&&&\\
\sze {${\it{P}}$ } & \sze 0.62 & \sze {\bf 0.47(1)$^\bullet$}& \sze 1.24 & \sze {$\times$} & \sze -- & \sze -- \\
\hline 
\end{tabular*}
\end{table}

\noindent
{\underline{Critical hopping parameter}}: The first observation that we would 
like to make, is that the value $\kappa_{crit}$ must be reached 
through a quadratic fit:
\bea
M_{PS}^2 &=& \alpha_1 (2 m_q) + \alpha_2 (2 m_q)^2
\label{e1}
\eea
where $2 m_q = \kappa^{-1} - \kappa_{crit}^{-1}$, is the quark mass 
that can be derived from the vector Ward identity (VWI). Namely, in order to be consistent with the ${\cal{O}}(a)$-improvement, 
the usual (linear) fit in quark mass should also be modified as $M_{PS}^2 \sim  \tilde{m}_q$, where $\tilde{m}_q = m_q ( 1 + b_m m_q)$. In perturbation theory, the coefficient $b_m < 0$. By fitting our data with~(\ref{e1}), we unambiguously obtain $\alpha_2 > 0$, which means that the positive curvature, due to the physical $m_q^2$ corrections, exceeds the lattice artifacts correction of the same order ($m_q^2$). The values of $\kappa_{crit}$ obtained from the quadratic and the linear fits differ appreciably ($\kappa_{crit}^{quad[lin]}=0.13585(2)[0.13576(2)]$). One more convincing argument in favor of $\kappa_{crit}^{quad}$, is provided by the ${\cal{O}}(a)$ improved axial Ward identity (AWI), by which the quark  mass is defined as:
\bea
\rho_q = {{\langle \partial_0 {A}_0(t) {\cal{O}}^{\dagger}(0)\rangle} \over { 2 \,\langle {P}(t) {\cal{O}}^{\dagger}(0)\rangle}}\, 
+\, {\cal{O}}(a^2).
\label{e2}
\eea
\noindent
From the linear fit $\rho_q(1/\kappa)$, we get $\lim_{\rho_q\to 0}{\kappa}=0.13584(5)$. A quadratic fit $\rho_q(1/\kappa)$ 
gives the same result and we deduce that $\kappa_{crit}^{quad}$ is {\sl the correct} value. 

\noindent
{\underline{Calibration of $a^{-1}$}}: We adopt the argument (of Ref.~\cite{allton}) that, due to the minimal extrapolation 
required, the most convenient quantity to fix $a^{-1}$ is $m_{K^*}$. We obtain: $a^{-1}=2.75(17)\, GeV$.
An encouraging feature of the improvement is that the value of the inverse lattice spacing, as obtained by using various 
physical quantities (e.g. $m_\rho$, $f_K$, $f_\pi$), remains stable and close to the one obtained from the computation of 
the string tension (the pure gluonic quantity). 

\noindent
{\underline{Light quark masses}}: 
The quark mass can be extracted in two ways, depending on the definition we chose, VWI or AWI. The two bare quark masses, 
when appropriately renormalized, should converge to the same physical result. Recently in Ref.~\cite{gimgim}, this was proved to be the case when the renormalization constants (RC) are computed non-perturbatively. We have also calculated the RC's (in the $RI$ scheme) by using the method for NP renormalization~\cite{mpstv} (see
Tab.~\ref{tab:tt1}). 
Subtleties and computational details on the extraction of bare lattice quark masses can be found in~\cite{qmass}. 
The $b$-coefficients, which are relevant for the improvement out of the chiral limit only, {\it i.e.} $Z_{S,P,A}(m,\mu)\simeq 
(1 + b_{S,P,A} ma) Z_{S,P,A}(0,\mu)$, are taken from the (boosted) perturbation theory. Our final results read:
\bea
m_{u,d}^{^{(\overline{\rm MS})}}(2\,GeV)&=&4.5(4)_{_{N^2LO}} [4.9(4)_{_{NLO}}]\,{\rm MeV},\cr
m_{s}^{^{(\overline{\rm MS})}}(2\,GeV)&=&111(12)_{_{N^2LO}} [121(13)_{_{NLO}}]\,{\rm MeV}.\nonumber 
\eea
Note that the matching between the $\overline{\rm MS}$ and $RI$ schemes was also performed at NNLO accuracy~\cite{vit}, which 
is the new result. To the same accuracy, the renorm. group invariant quark masses are:
 \\
$\hat{m}_{u,d} =7.2(6)\,{\rm MeV},\quad \hat{m}_{s} =177(19)\,{\rm MeV}$.\\
\vspace*{-2.5mm}

\noindent
{\underline{Light meson decay constants}}: 
For the full ${\cal{O}}(a)$ improvement of the quark bilinear operators, the coefficients $c_J$, of the appropriate 
counterterms, are tuned to their non-perturbative values. Only $c_{T}$ is to be taken from the  boosted perturbation theory. 
By using the improved currents matrix elements, we define decay constants in the following way:
\bea
\langle 0\vert A_0\vert{\cal{P}}\rangle &=& i M_{{\cal{P}}} \,\left[ {F}_{{\cal{P}}}^{(0)}+ c_A a {F}_{{\cal{P}}}^{(1)} \right], \cr
\langle 0\vert V_i\vert{\cal{V}}^{(\lambda)}\rangle &=& i e_i^{(\lambda)} M_{{\cal{V}}} \,\left[ {F}_{{\cal{V}}}^{(0)}+ c_V a {F}_{{\cal{V}}}^{(1)} \right], 
\cr
\langle 0\vert T_{i0}\vert{\cal{V}}^{(\lambda)}\rangle &=& i e_i^{(\lambda)} M_{{\cal{V}}} \,\left[ {F}_{{\cal{V}}}^{T(0)}+ c_V a {F}_{{\cal{V}}}^{T(1)} \right] 
\label{e3}
\eea
where ${\cal{P}}$, $\cal{V}$ denote the pseudoscalar and vector mesons at rest. The corrections $a F_{\cal{J}}^{(1)}$, 
correspond to the matrix element of: the gradient of the pseudoscalar density, the divergence of tensor current, and to the 
curl of the vector current, respectively. Empirically, $a F_{\cal{P}}^{(1)}$ is large, whereas $a F_{\cal{V}}^{(1)}$ and 
$a F_{\cal{V}}^{T(1)}$ are small. However, when multiplied by corresponding $c_J$, their respective contributions never 
exceed $\sim 5\%$, relative to the leading term ($F_{\cal{J}}^{(0)}$). 

All the bare lattice local currents in (\ref{e3}), should be multiplied by appropriate renormalization constants. In the 
chiral limit, they are all computed nonperturbatively (see Tab.~\ref{tab:tt1}). This is not the case for $b_J$'s, which 
are all (except for $b_V$) taken from 1-$loop$ boosted perturbation theory. 
With all these ingredients, we calculated the pseudoscalar and vector decay constants. The results, obtained using 
$a^{-1}(m_{K^*})$, are listed in Tab.~2.
\begin{table}[h!]
\setlength{\tabcolsep}{0.25pc}
\caption{\sl \small Decay constants.
\vspace*{1mm}}
\label{tab:tt2}
\hspace*{-2mm}
\vspace*{-6mm}
\begin{tabular*}
{7.225cm}{ccc}
\hline
\vspace*{-3.5mm}&&\\
{$f_\pi$ } & {$f_K$ } & {$f_K/f_\pi$ } \\
\vspace*{-3mm}&&\\
{$139\pm 22\,{\rm MeV}$ } & {$156\pm 17\,{\rm MeV}$ } & {$1.13\pm 0.04$ } \\ \hline
\vspace*{-3.5mm}&&\\
{$f_\rho$ } & {$f_{K^*}$ } & {$f_\phi$ } \\
\vspace*{-4mm}&&\\
{$199\pm 15\,{\rm MeV}$ } & {$219\pm 7\,{\rm MeV}$ } & {$235\pm 4\,{\rm MeV}$ }\\ \hline 
\vspace*{-3.5mm}&&\\
{$f_\rho^T (2\,{\rm GeV})$ } & {$f_{K^*}^T (2\,{\rm GeV})$ } & {$f_\phi^T (2\,{\rm GeV})$ } \\
\vspace*{-3.5mm}&&\\
{$165\pm 11\,{\rm MeV}$ } & {$178\pm 10\,{\rm MeV}$ } & {$212\pm 7\,{\rm MeV}$ } \\ \hline
\end{tabular*}
\end{table}

\vspace*{-1mm}
\noindent
It is well known that the ratio 
$f_K/f_\pi - 1$ (which is the `measure' of the SU(3) flavor breaking), as extracted from the lattice calculation, 
is always smaller than the experimental value ($0.22$). After a careful analysis of all the data (unimproved and improved), 
the lesson we draw is: {\sl $(f_K - f_\pi)/f_\pi$ from the lattice quenched studies is insensitive to the improvement and 
remains more than $30\%$ lower than its experimental value}.\\
\noindent
On the other hand, when compared to the unimproved results, the values for the vector decay constants are in substantially
better agreement with the experimental ones. For comparison with the unimproved results, we show the directly extracted data 
(with {\bf no} extrapolations), as obtained from simulations with unimproved, tree-level improved and NP improved Wilson fermions. 
\begin{figure}[hb]
\vspace*{-.65cm}
\hspace*{-.25cm}
\epsfig{file=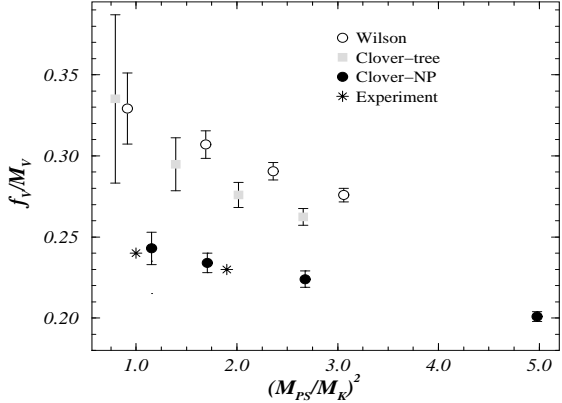,width=7.5cm} 
\vspace*{-1.cm}
\caption{\footnotesize Comparison of directly extracted vector meson decay constant with unimproved, tree-level improved 
and NP improved Wilson fermions, at the same $\beta = 6.2$, and on the same lattice, $24^2\times 64$.}  
\label{fig.1}  
\vspace*{-.8cm}
\end{figure}
The other source of vector mesons is provided by the tensor current. The masses extracted from vector and tensor correlation 
functions are (of course) the same, while the couplings are different. In Tab.~2, we give the first lattice predictions for 
$f^{T}_{\cal{\rho ,{\sl K}^*\! \! , \phi}}(2\,{\rm GeV})$.

\vspace*{.1cm}

\noindent
{\underline{Heavy-light meson decay constant}} ({\it preliminary results}):
In the simulations, we included four heavy quarks which we combined with the four light ones. After a quadratic 
extrapolation/interpolation in light quark masses, we interpolate in the charm region to estimate $f_D$ and $f_{D_s}$. 
This results in: 
\bea
f_D &=& 202 \pm 14\,{}^{+0}_{-12}\, {\rm MeV},\nonumber \\
f_{D_s} &=& 231 \pm 11\,{}^{+7}_{-0} \, {\rm MeV},\nonumber \\
{f_{D_s}/f_{D}} &=& 1.11\pm 0.03.  
\eea
The stability of this result ({\it see} \cite{prepa}), does not hold for $B$ mesons. As usual, we have to 
extrapolate our results by using the heavy quark scaling law:
\bea
f_{Qq} = {{\frak{a}}_0 \over \sqrt{m_{Qq}}}\left( 1 +  {{\frak{a}}_1 \over m_{Qq}}+  {{\frak{a}}_2 \over m_{Qq}^2}\right).
\eea
 It turns out that $f_{B,B_s}$ are sensitive to whether we include the $\sim 1/m_{Qq}^2$ term in the fit or not. This is 
 illustrated in Fig.~2, and its effects, along with the scale fixing dependence, are included in the systematic error. Our results are:
\bea
f_B\,&=&\,177 \pm 19\,{}^{+29}_{-10}\,{\rm MeV}, \nonumber \\
f_{B_s}\,&=&\,206 \pm 16\,{}^{+32}_{-0}\,{\rm MeV}, \nonumber \\
{f_{B_s}/ f_B}\,&=&\,1.16\pm 0.04. 
\eea
\begin{figure}[h!]
\vspace*{-1.25cm}
\epsfig{file=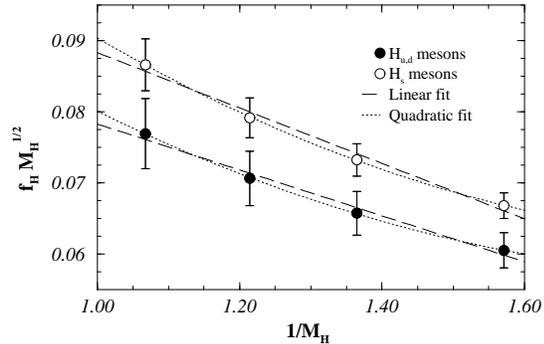,width=7.5cm} 
\vspace*{-1.5cm}
\caption{\footnotesize Directly extracted {\it heavy-light} pseudoscalar decay constants vs. two forms of fit needed to 
extrapolate to the B meson sector.}  
\label{fig.2}  
\end{figure}
\vspace*{-1.2cm}


\end{document}